\documentclass[10pt]{article}
\usepackage{graphicx}

\def\Title#1{\begin{center} {\Large #1 } \end{center}}
\def\Author#1{\begin{center}{ \sc #1} \end{center}}
\def\Address#1{\begin{center}{ \it #1} \end{center}}

\newcommand\pubblock{\rightline{\begin{tabular}{l} Proceedings of the Fifth Annual LHCP\\ \pubnumber\\
         \pubdate  \end{tabular}}}

\newenvironment{Abstract}{\begin{quotation} \begin{center} 
             \large ABSTRACT \end{center}\bigskip 
      \begin{center}\begin{large}}{\end{large}\end{center} \end{quotation}}

\newenvironment{Presented}{\begin{quotation} \begin{center} 
             PRESENTED AT\end{center}\bigskip 
      \begin{center}\begin{large}}{\end{large}\end{center} \end{quotation}}

\def\Acknowledgements{\bigskip  \bigskip \begin{center} \begin{large}
             \bf ACKNOWLEDGEMENTS \end{large}\end{center}}

\def\beq{\begin{equation}}
\def\eeq#1{\label{#1}\end{equation}}
\def\eeqn{\end{equation}}

\def\beqa{\begin{eqnarray}}
\def\eeqa#1{\label{#1}\end{eqnarray}}
\def\eeqan{\end{eqnarray}}

\let\bar=\overbar

\def\Dslash{\not{\hbox{\kern-4pt $D$}}}
\def\dslash{\not{\hbox{\kern-2pt $\del$}}}

\def\msb{{\bar{\ssstyle M \kern -1pt S}}}

\usepackage{color}
\usepackage{cite}

\newcommand\pubnumber{ ATL-PHYS-PROC-2017-125 }

\newcommand\pubdate{\today}

\def\affiliation{
On behalf of the ATLAS Collaboration, \\
Physics Division, Lawrence Berkeley National Laboratory\\
Berkeley, CA 94720, USA}

\begin{document}

\large
\begin{titlepage}
\pubblock

\vfill
\Title{  Jet and photon measurements with ATLAS }
\vfill

\Author{ Benjamin Nachman  }
\Address{\affiliation}
\vfill
\begin{Abstract}

The strong force is responsible for a rich set of phenomena that can be probed using a variety of techniques over a wide energy and angular range at the Large Hadron Collider.  This talk reports on the latest results from the ATLAS Collaboration that measure the high energy, wide angle, collinear, and soft regimes of quantum chromodynamics (QCD).  There is also an important connection between QCD at high energies and electroweak phenomena including massive gauge bosons as well as photons.  

\end{Abstract}
\vfill

\begin{Presented}
The Fifth Annual Conference\\
 on Large Hadron Collider Physics \\
Shanghai Jiao Tong University, Shanghai, China\\ 
May 15-20, 2017
\end{Presented}
\vfill
\end{titlepage}
\def\thefootnote{\fnsymbol{footnote}}
\setcounter{footnote}{0}
%

\normalsize 


\section{Introduction}

\begin{figure}[htb]
\centering
\includegraphics[width=0.5\textwidth]{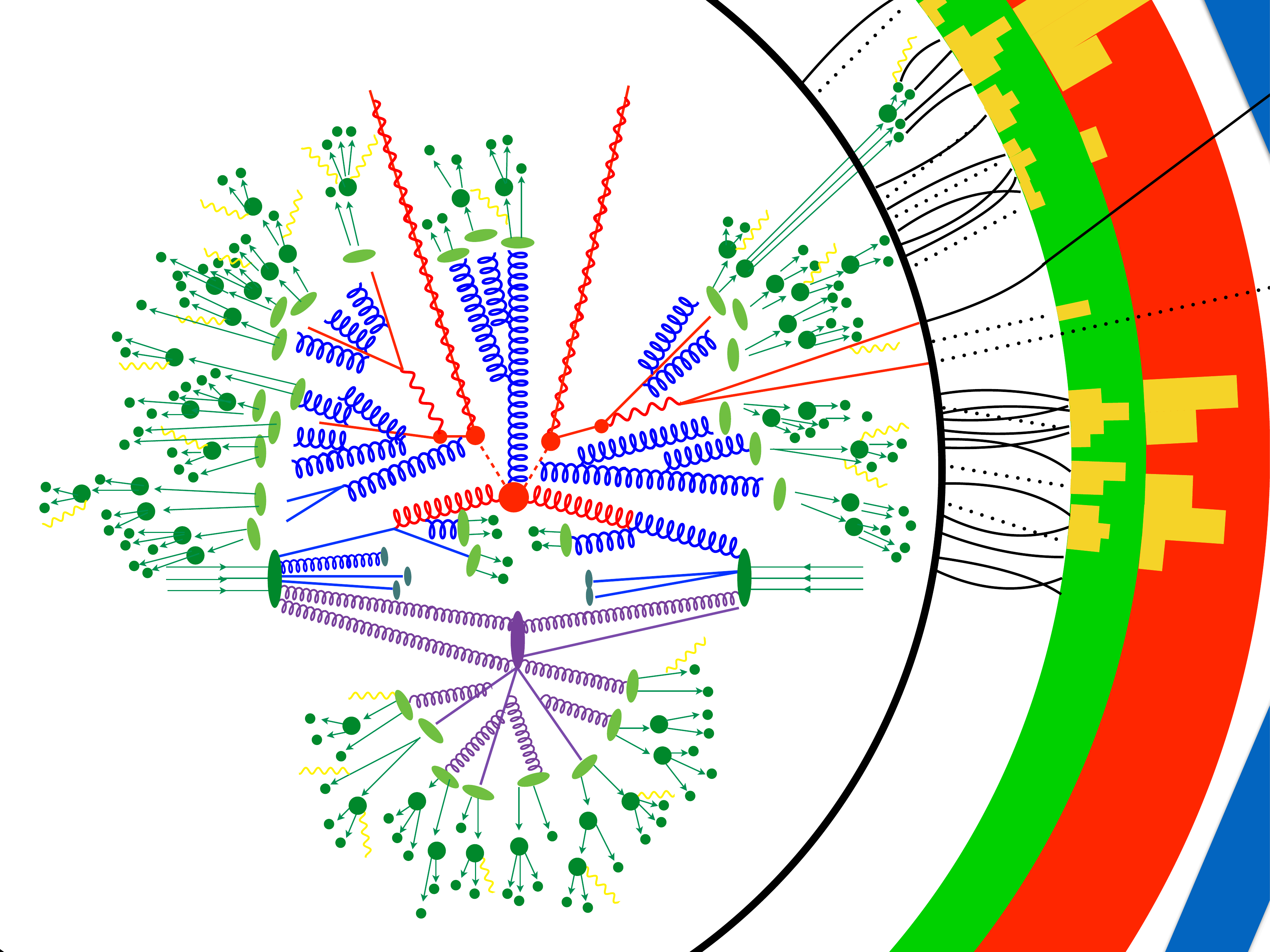}
\caption{A schematic diagram illustrating the complex nature of the strong force and its consequences spanning $10^{-19}$ m up to $10^1$ m. Not to scale.  Modeled after (but not the same as) Fig. 1 in Ref.~\cite{Gleisberg:2008ta}.}
\label{fig:figure1}
\end{figure}

Quantum chromodynamics (QCD) is responsible for most of the phenomena at the LHC over a wide range of energy and angular scales - see Fig.~\ref{fig:figure1}.  The observable consequence of strong force carrying particles is jets: collimated sprays of particles resulting from quarks and gluons produced at high energy.  The energy spectrum of jets, the angular distribution of jets, and the radiation pattern inside jets are all\footnote{Of course, $\alpha_s$ is really a function that depends on the energy scaled probed; also, there are the number of particles in the various representations of SU(3) that is in principle a `free' parameter of QCD.} governed by a single number: $\alpha_s$.  Measurements at the LHC can probe regions of QCD phase space that are well-described by fixed order calculations (well-separated jets), the resummation regime (soft and collinear physics), as well as the non-perturbative regime.  The following sections highlight measurements using the ATLAS detector~\cite{atlas} that are sensitive to each of these regimes and allow us to study how jets emerge from the underlying theory.

\section{Jet Multiplicities and Energies}

\begin{figure}[htb]
\centering
\includegraphics[width=0.8\textwidth]{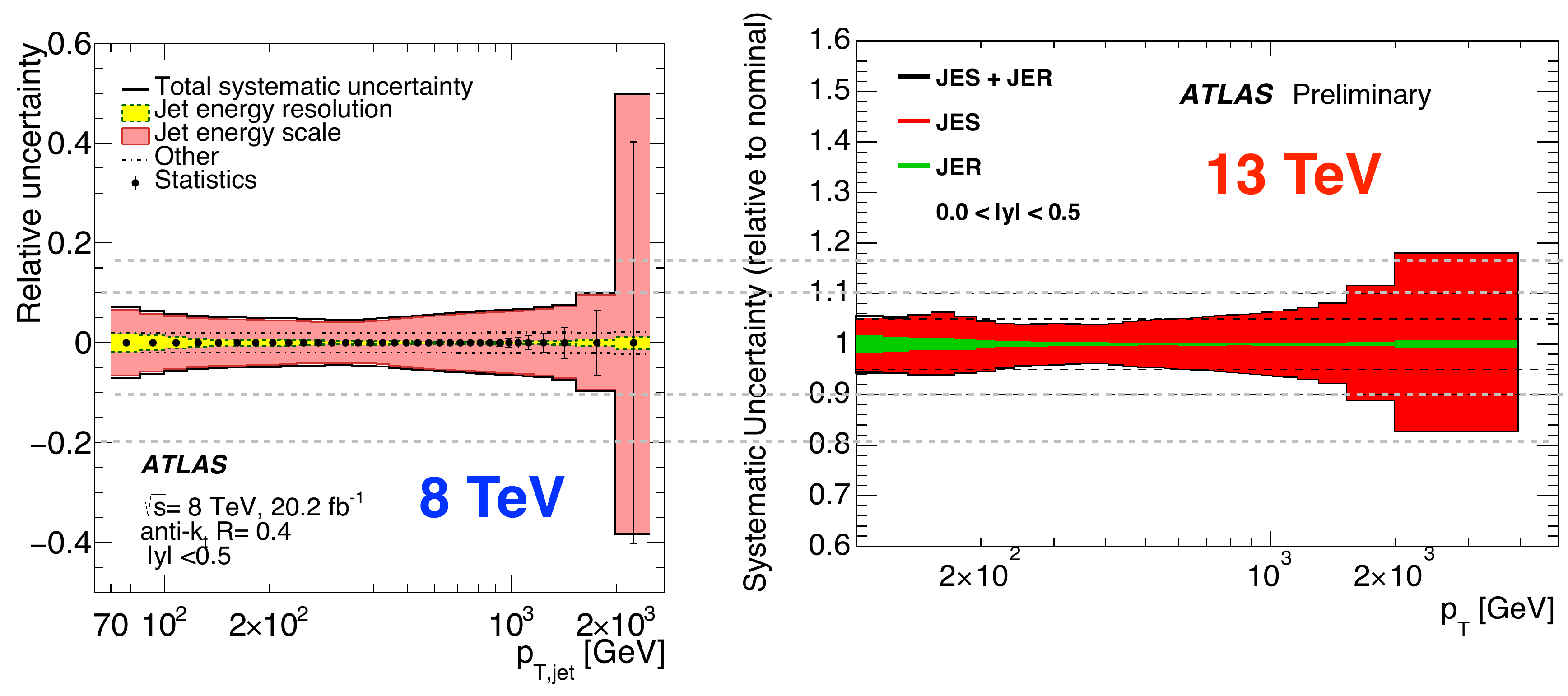}
\caption{ Jet cross section uncertainty from $\sqrt{s}=8$ TeV~\cite{jet1} (right) and $\sqrt{s}=13$ TeV~\cite{jet2} (right).}
\label{fig:figure2}
\end{figure}

Jet cross sections measured over a range in rapidities and energies accessible at the LHC span many orders of magnitude.  With only 3.2 fb${}^{-1}$, the $\sqrt{s}=13$ TeV Run 2 inclusive jet cross section measurement is able to probe higher jet energies ($\sim 3 $ TeV) than was possible with the full Run 1 dataset ($\sim 2 $ TeV).  The cross section measurement uncertainty is dominated by the precision of the jet energy scale (JES) and already in early Run 2, the JES uncertainty is competitive with Run 1 (Fig.~\ref{fig:figure2}).  These data provide a stringent test of the most precise fixed-order pure QCD calculations currently available as well as constraining parton distribution functions (PDFs) in the high momentum fraction regime.

\begin{figure}[htb]
\centering
\includegraphics[width=0.8\textwidth]{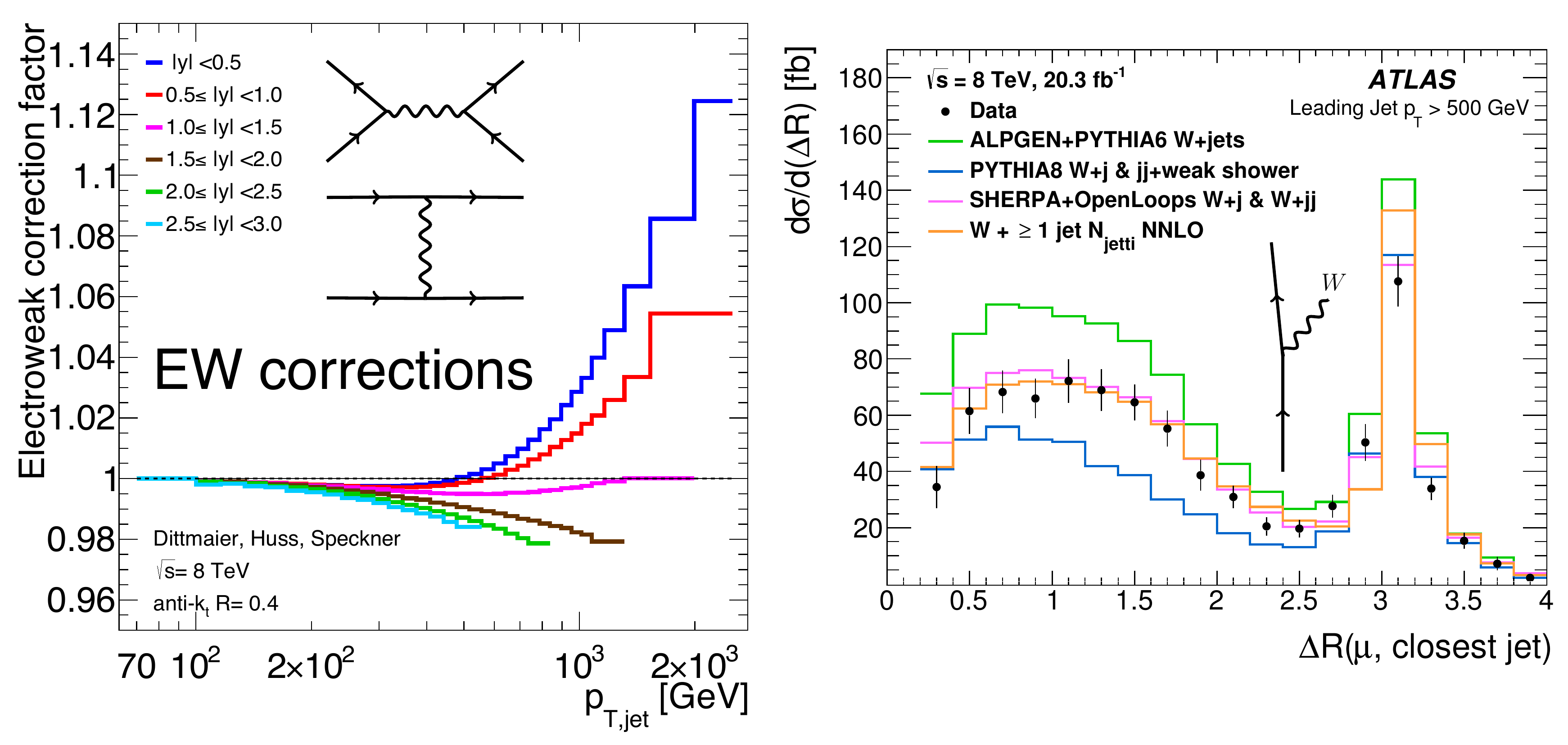}
\caption{ The size of electroweak corrections~\cite{jet1,ewcor} (left) as well as a recent measurement of real electroweak boson radiation~\cite{realW,realWNNLO,realWalpgen,pythia8,sherpa1,sherpa2} (right).}
\label{fig:figure3}
\end{figure}

The theoretical interpretation of the cross section at very high energies is conceptually cleaner than at low $p_\mathrm{T}$ because non-perturbative effects from hadronization are  negligible.  However, in the multi-TeV regime, electroweak corrections are not small.  The left plot of Fig.~\ref{fig:figure3} shows that at $p_\mathrm{T}\sim 1$ TeV, electroweak corrections can reach $\sim 10\%$, which exceeds the uncertainty on the measurement.  It is therefore critical for accurate predictions that these contributions to the cross section are included.  In fact, real electroweak boson emission is large enough to be measurable, as indicated by the right plot of Fig.~\ref{fig:figure3}.  At low energies, the radiation of a real $W$ or $Z$ boson is severely suppressed by their masses, but this becomes increasingly irrelevant at high energies.   This real boson emission is not well-modeled by all modern simulation setups, including those including dedicated electroweak radiation in the parton shower.

\begin{figure}[htb]
\centering
\includegraphics[width=0.8\textwidth]{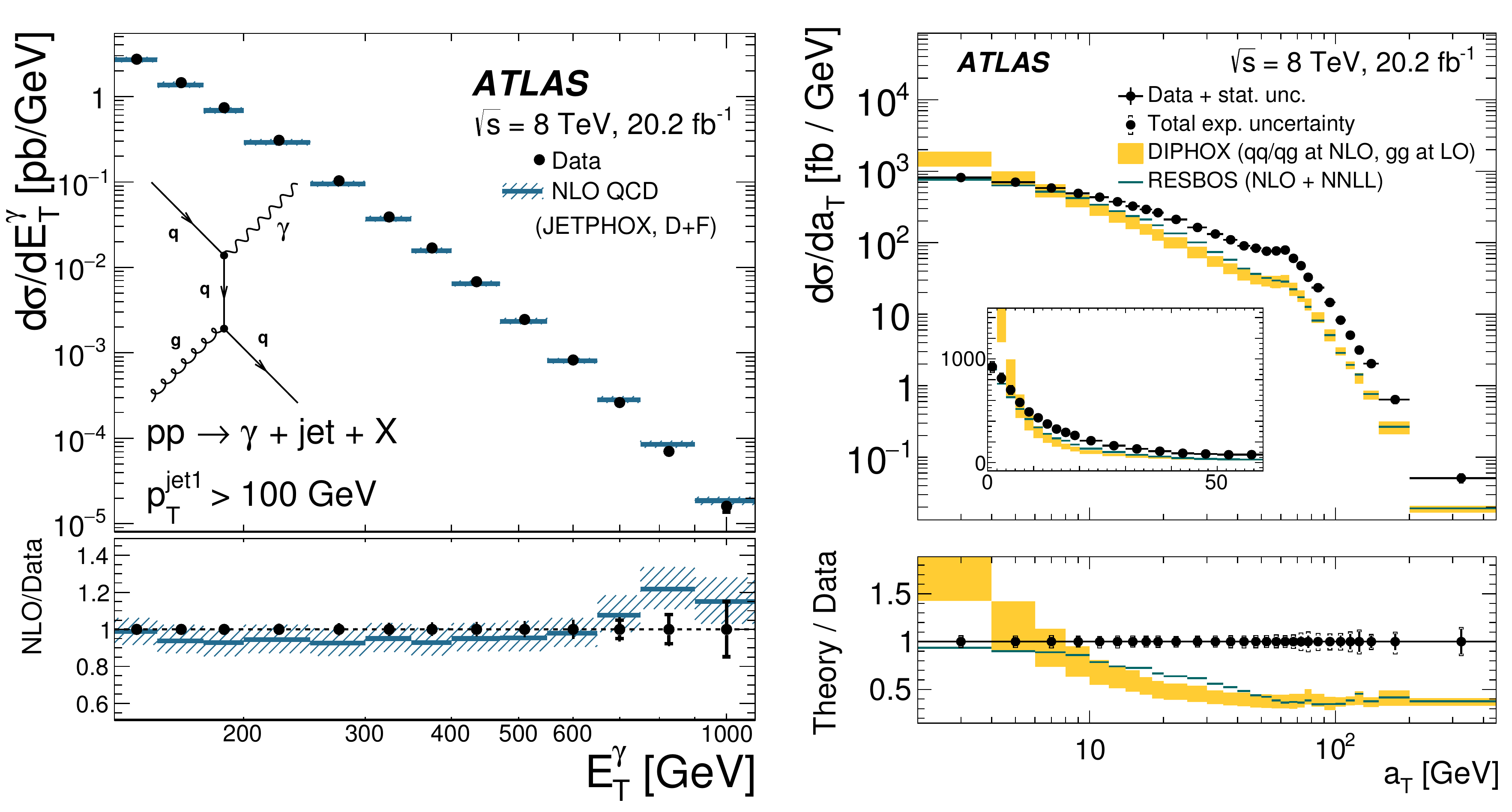}
\caption{ The energy spectrum of inclusive photon production~\cite{photon3} (left) and the $p_\mathrm{T}$ perpendicular to the transverse thrust axis ($a_\mathrm{T}$) in diphoton events~\cite{photon2,photon2diphox,photon2resbos1,photon2resbos2,photon2resbos3} (right).  The inset in right plot shows the low momentum regime that is particularly sensitive to soft gluon emission and is well-modeled by resummation.}
\label{fig:figurephoton}
\end{figure}

For a wide range of energies, real photon emission also offers the opportunity for a precision study of the interplay between electroweak and QCD phenomena.  The ATLAS collaboration has recently performed a series of inclusive and exclusive (multi)photon measurements both at $
\sqrt{s}=8$ and $\sqrt{s}=13$ TeV~\cite{photon1,photon2,photon3}.  Inclusive, well-isolated photons are well-described by next-to-leading order (NLO) QCD (left plot of Fig.~\ref{fig:figurephoton}), though this is not the case for photon pairs (right plot of Fig.~\ref{fig:figurephoton}).  Resummation is important to improve the modeling in the region of low transverse region - a theme that will reoccur also in the collinear regime for both event shapes and jet shapes in the sections that follow.

\section{Event Shapes}

\begin{figure}[htb]
\centering
\includegraphics[width=0.8\textwidth]{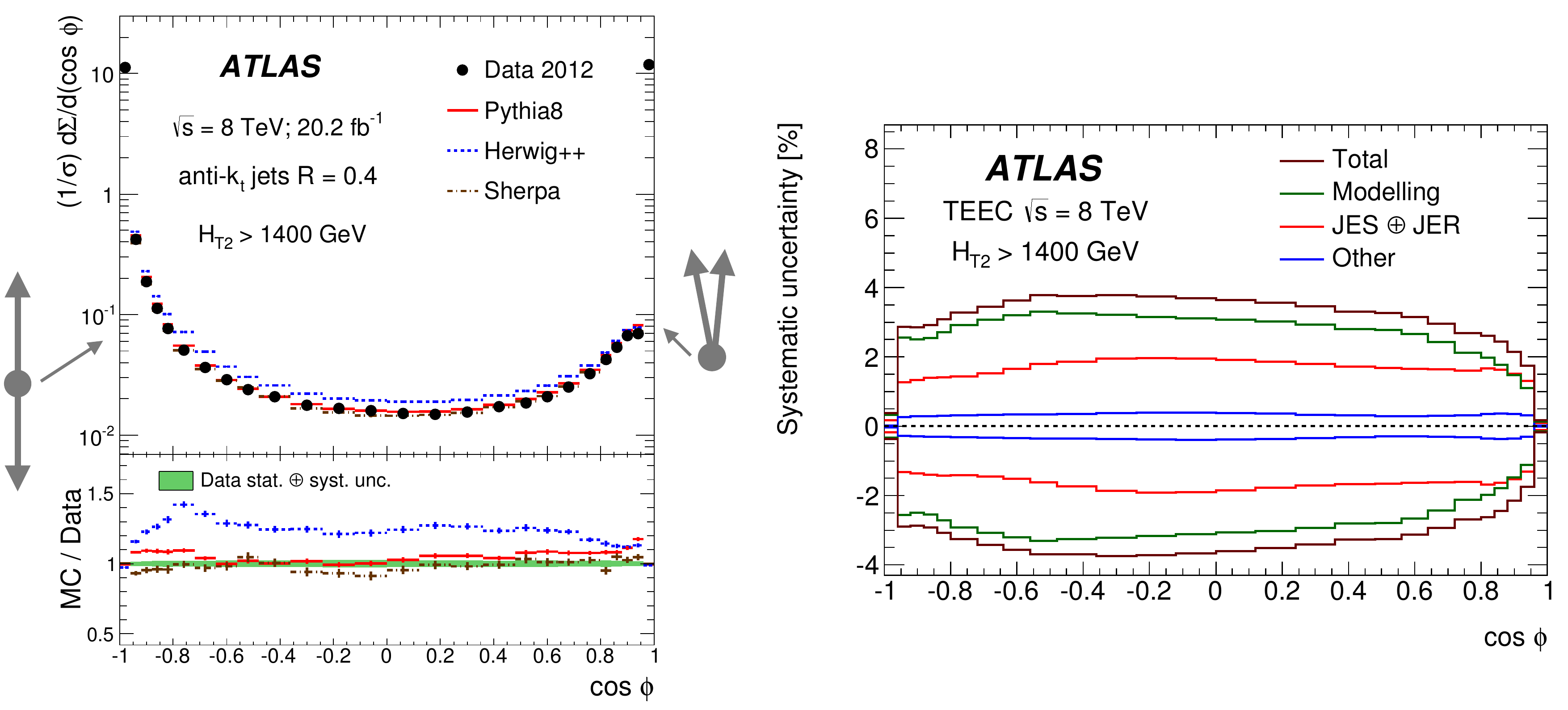}
\caption{Left: The distribution of the Transverse Energy Energy Correlation (TEEC) as a function of the cosine of the azimuthal opening angle between jet pairs; schematic diagrams to the left and right of the figure illustrate representative jet topologies.  Right: a breakdown of the systematic uncertainty on the TEEC~\cite{eventshapes}.}
\label{fig:figure4}
\end{figure}

Event shapes like thrust~\cite{thrust} have played a key role in the precision QCD program at $e^+e^-$ colliders in the past (see e.g. Ref.~\cite{qcdatepem}).  Modifications of these observables are also powerful tools at a hadron collider because they tend to be dimensionless ratios that are largely insensitive to systematic uncertainties on the JES.  One observable that has recently been measured with the full Run 1 dataset is the transverse energy-energy correlation function (TEEC)~\cite{eventshapes}.  The TEEC is the product of transverse momenta for pairs of jets normalized by the total transverse momentum in the event.  To expose different (multi)jet configurations, the TEEC is measured as a function of the transverse angle between the jet pairs (Fig.~\ref{fig:figure4}).  To probe different scales, the measurement is also performed as a function of the sum of the transverse momenta of the leading two jets, $H_\mathrm{T2}$.  Parton shower Monte Carlo (MC) provides an excellent model of the TEEC distribution even for $\cos\phi\sim \pm 1$ where soft and collinear physics are important, respectively, and fixed order QCD is unable to describe the data.  As advertised, the JES uncertainty is significantly reduced with respect to the jet cross section measurement, as indicated by the right plot of Fig.~\ref{fig:figure4}.

The availability of precise fixed order QCD calculations (LO $= \mathcal{O}(\alpha_s^3)$; NLO is $ \mathcal{O}(\alpha_s^4)$) alongside the precise measurement, allows for the extraction of $\alpha_s$.  Without resummation, the region near $\cos\phi\sim \pm 1$ must be removed from the fiducial phase space.   The resulting extraction is a useful input to the study of $\alpha_s$, as it provides a large lever arm for measuring the running of the coupling.  The experimental uncertainty from this extraction is competitive with other determinations of $\alpha_s$ in ATLAS and CMS, though currently scale uncertainties are dominant and are the limiting factor for increasing the precision.

\section{Inside Jets}

The event shapes measurement from the previous section tried to avoid the soft and collinear limits of QCD; jet shapes - measurements of the radiation inside jets - are dominated by these phenomena.   Significant theoretical and experimental advances in recent years have resulted in a burst of activity in the area of precision jet substructure (see e.g. Ref.~\cite{boostreview} and the many papers that cite it).  This section highlights an innovative use of a $\sqrt{s}=8$ TeV precision jet substructure measurement to derive uncertainties for a jet tagger in Run 2~\cite{pub}.

\begin{figure}[htb]
\centering
\includegraphics[width=0.5\textwidth]{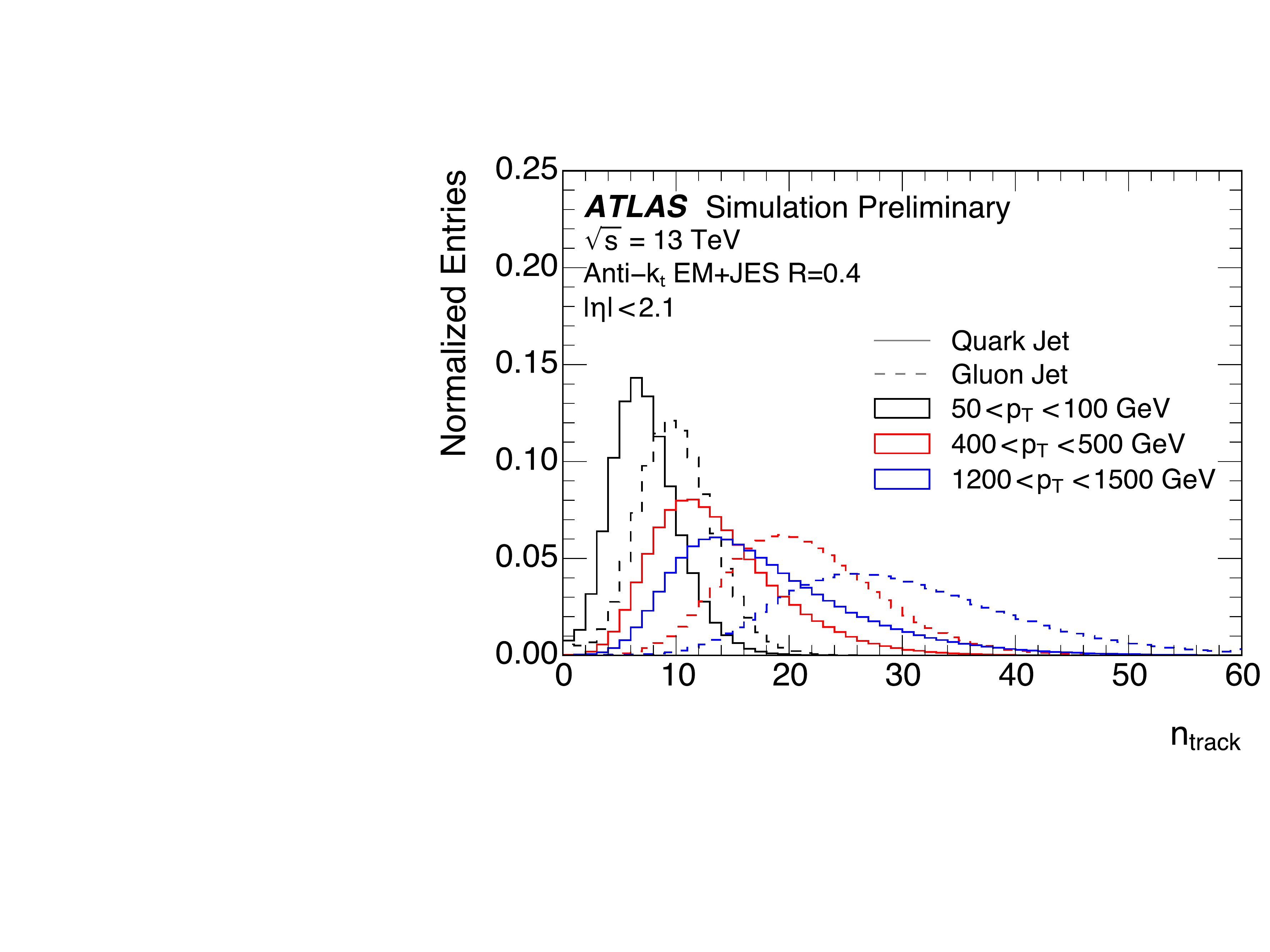}
\caption{ The number of tracks reconstructed inside quark and gluon jets in three bins of jet $p_\mathrm{T}$ using the Pythia 8 generator~\cite{pub,pythia8}.}
\label{fig:figure5}
\end{figure}

The number of particles inside a jet is one of the most powerful tool for distinguishing quark-iniated (`quark') from gluon-initiated (`gluon') jets.  The number of reconstructed tracks is an observable related to the total number of particles and has a long history for quark/gluon tagging.  Figure~\ref{fig:figure5} shows the separation between quark and gluon jets and how the track multiplicity ($n_\mathrm{track}$) increases with jet $p_\mathrm{T}$.  The uncertainty in the $n_\mathrm{track}$ distribution can be decomposed into a contribution from the charged particle multiplicity and a second contribution from the reconstruction of tracks from charged particles.  The former  is constrained using the $\sqrt{s}=8$ TeV measurement of the quark/gluon charged particle multiplicity and the latter is covered using track reconstruction uncertainties developed with the Run 2 detector.  This approach is valid because even though the reconstruction efficiency of charged particles as tracks changed between Runs 1 and 2, the particle-level multiplicity distribution did not.

\begin{figure}[htb]
\centering
\includegraphics[width=0.7\textwidth]{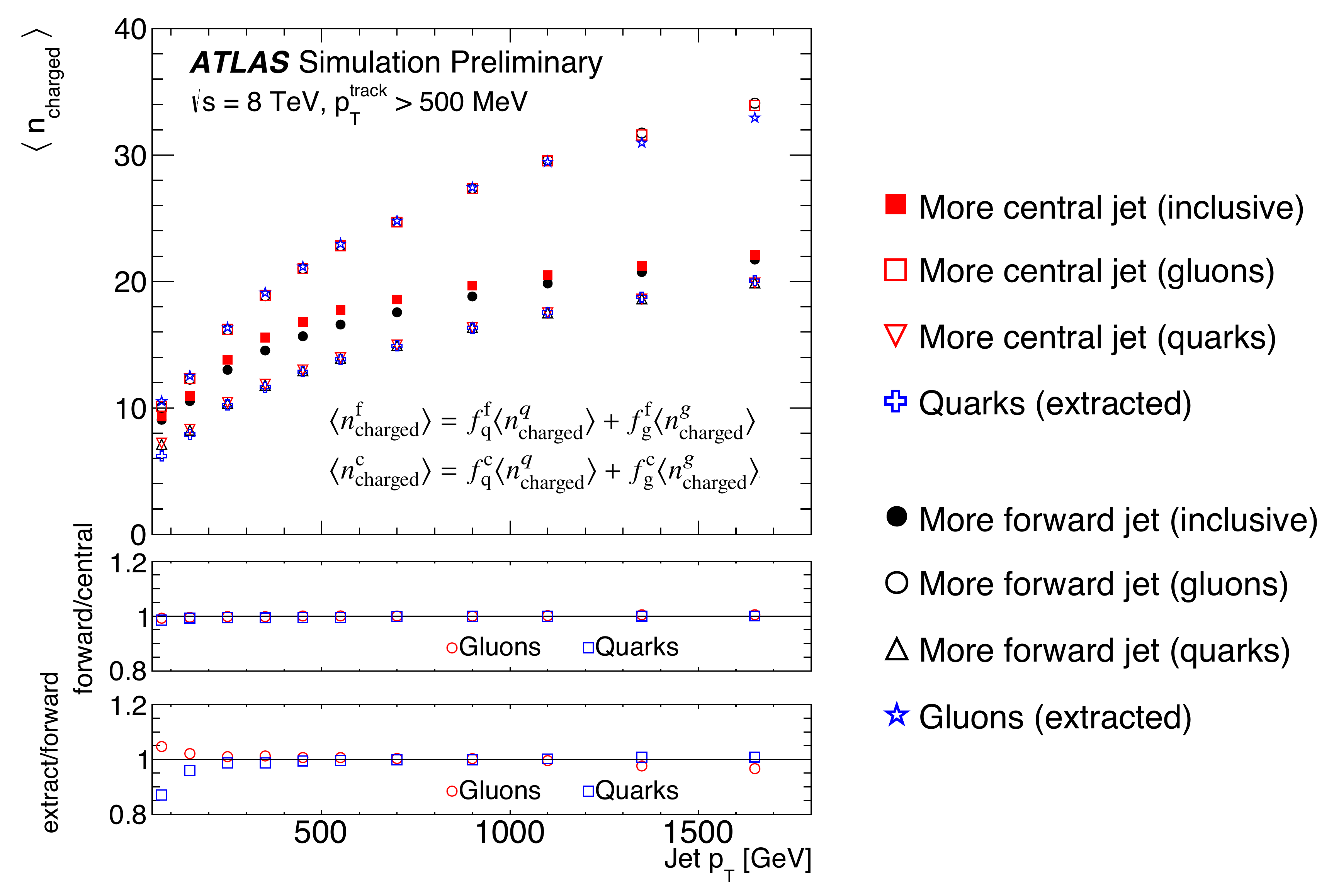}
\caption{ A demonstration of the extraction and closure of the method for determining the quark/gluon charged particle multiplicity.  The equations in the upper plot are used for turning the measurement of the forward/central jet charged particle multiplicity into the quark/gluon charged particle multiplicity given the input quark/gluon fractions $f$~\cite{paper,pub}.}
\label{fig:figure6}
\end{figure}

\begin{figure}[htb]
\centering
\includegraphics[width=0.95\textwidth]{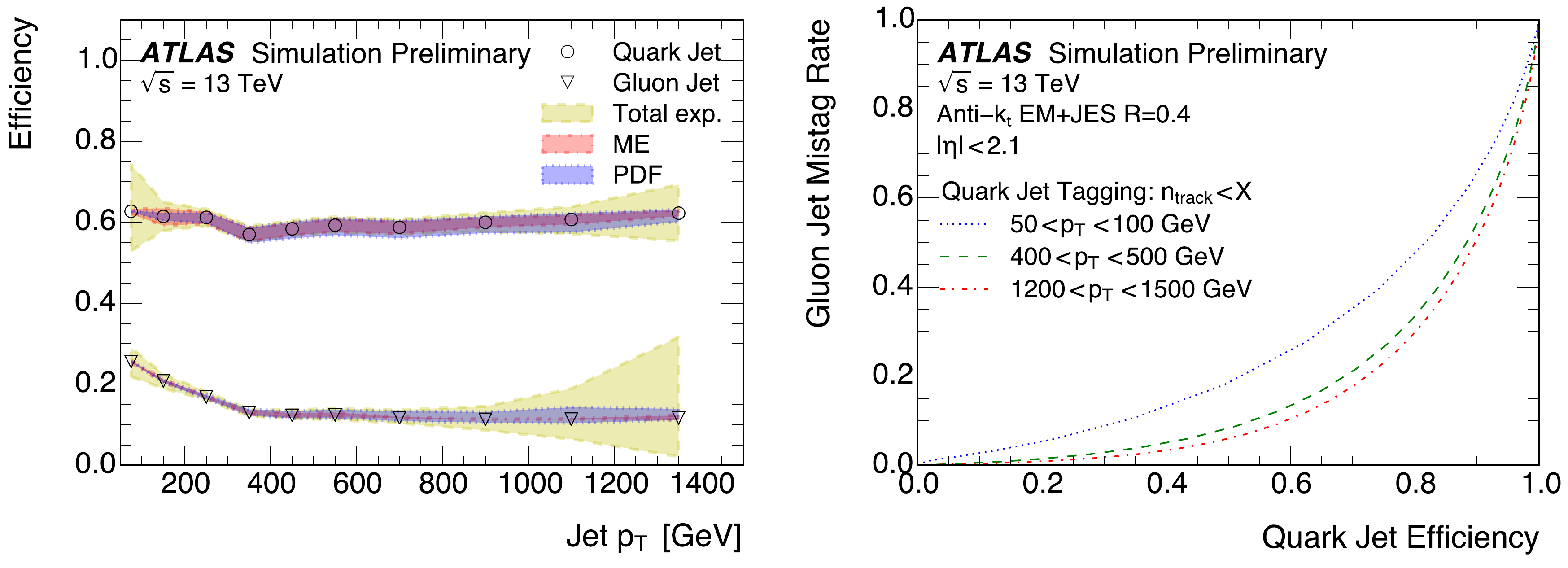}
\caption{ Left: for a 60\% quark jet efficiency, plotted is the quark/gluon jet efficiency along with the associated systematic uncertainty.  Right: the performance of the charged particle multiplicity based quark/gluon tagger in three jet $p_\mathrm{T}$ bins~\cite{pub}.}
\label{fig:figure7}
\end{figure}

The extraction of the charged particle multiplicity for quark and gluon jets separately exploits the fact that in a well-balanced dijet event, the more forward of the two jets is more likely to be quark jet.  Given the fraction of quark and gluon jets for a given jet $p_\mathrm{T}$, the system of equations shown in Fig.~\ref{fig:figure6} is solved for $n_\mathrm{charged}^q$ and $n_\mathrm{charged}^g$.  This extraction procedure assumes that the multiplicity only depends on the jet $p_\mathrm{T}$ and type, which is validated across a wide range in jet $p_\mathrm{T}$ in Fig.~\ref{fig:figure6}.  The uncertainty on the measurement is then combined with uncertainties on the quark/gluon fractions to arrive at a complete particle-level uncertainty on the quark/gluon tagger performance, shown in Fig.~\ref{fig:figure7}.  The total uncertainty is $\sim 5\%$ over a wide range of jet $p_\mathrm{T}$ and the tagger achieves a gluon jet rejection of about 10 for a 50\% quark jet efficiency.  This technique may be amenable to more sophisticated jet substructure observables and provides a robust alternative to traditional methods of comparing jets between different samples such as dijets and $Z$+jets. 

\section{Conclusions / Outlook}

The ATLAS collaboration has an active program for measuring jets and related phenomena across all accessible regions.  These studies are important for testing QCD in a variety of unique ways (fixed order, resummation, non-perturbative regimes), for developing tagging techniques that may be useful for searches for other Standard Model (SM) and beyond the SM physics analyses, and for tuning free parameters of our current models for the best possible prescription of the data.  While this manuscript was in preparation, one of the analyses described above (Ref.~\cite{paper}) has already played an important role in improving the quark/gluon jet description in the state-of-the-art Herwig 7 MC generator~\cite{herwig71}.  This is only the beginning of a rich and hopefully productive program that will continue to benefit from close connections between the experimental and theoretical communities. 

\Acknowledgements
This work is supported by the DOE under contract DE-AC02-05CH11231.

\end{document}